# Estimation Large- Scale Fading Channels for Transmit Orthogonal Pilot Reuse Sequences in Massive MIMO System

Qazwan Abdullah
community college of Yareem
Ibb, Yemen
gazwan20062015@gmail.com

Nor Shahida Mohd Shah
Faculty of Faculty of Engineering Technology, Universiti Tun Hussein Onn Malaysia, Pagoh, Muar, Johor, Malaysia.
shahida@uthm.edu.my

Shipun Hamzah
Faculty of Electrical and Electronic Engineering
Universiti Tun Hussein Onn Malaysia
Johor, Malaysia
shipun@uthm..edu.my

Adeb Salh
community college of Yareem
Ibb, Yemen
adebali@utm.edu.my

Mahathir Mohamad
Faculty of Applied Sciences and Technology, Universiti Tun Hussein Onn Malaysia, Pagoh, Muar, Johor, Malaysia.
mahatir@uthm..edu.my

Shahilah Nordin
Faculty of Electrical Engineering
Universiti Teknologi Mara(UiTM),Malaysia
Pulau Pinang, Malaysia
shahilah879@ppinang.uitm.edu.my

Maisarah Abu
Faculty of Electronic and Computer Engineering, University Technical Malaysia Malacca
Malacca City, Malacca
maisarah@utem.edu.my

Mohammed Abdo Albaom
Department of Computer Science, Faculty of Computer Science and Information Technology Universiti Putra
Serdang, Malaysia
 mohammedabdo773@gmail.com

safwan sadeq
Department of Mechanical Engineering, Universiti Teknologi Petronas
Perak, Malaysia
asafwanalshahri@yahoo.com

*Abstract*— Massive multiple-input multiple-output (MIMO) is a critical technology for future fifth-generation (5G) systems. Reduced pilot contamination (PC) enhanced system performance, and reduced inter-cell interference and improved channel estimation. However, because the pilot sequence transmitted by users in a single cell to neighboring cells is not orthogonal, massive MIMO systems are still constrained. We propose channel evaluation using orthogonal pilot reuse sequences (PRS) and zero forced (ZF) pre-coding techniques to eliminate channel quality in end users with poor channel quality based on channel evaluation, large-scale shutdown evaluation, and analysis of maximum transmission efficiency. We derived the lower bounds on the downlink data rate (DR) and signal-to-interference noise ratio (SINR) that can be achieved based on PRS assignment to a group of users where the number of antenna elements mitigated the interference when the number of antennas reaches infinity. The channel coherence interval limitation, the orthogonal PRS cannot be allocated to all UEs in each cell. The short coherence intervals able to reduce the PC and improve the quality of channel. The results of the modelling showed that higher DR can be achieved due to better channel evaluation and lower loss.

## I. INTRODUCTION

Massive multi-input multiple-output (MIMO) is a critical technology to reduce interference between adjacent cells and multi-user interference to increase the capacity of this 5G network. Massive MIMO technologies have recently received much attention as a promising technology to significantly increase spectrum efficiency. In multi-cell massive MIMO systems, pilot contamination (PC) is a essential problem, which influences the Data Rate (DR). Nevertheless, training channel and reducing the interference depend on mitigated pilot reuse sequences (PRS), where the PRS increase gain and avoid interference between neighboring cells [1-5]. The pilot scheduling takes into account how the system allocates the pilot sequences to users in order to minimize or even simply eliminate the PC problems. The PC is a major problem in multicellular massive MIMO systems, and it affects DR. However, channel evaluation can be used to train the channel by conducting PRS in a time distribution duplex (TDD), which increases the benefit of PRS between neighboring cells and eliminates interference between neighboring cells [1-5]. Placing a very large number of antennas on a transmitter, linear signal processing can significantly reduce the PC from any rapid fading. The large-scale fading cannot be ignored, which are a crucial technique for estimating system performance, and evaluation of the feasible data rate based on the effective signal-to-interference noise ratio (SINR). The received training signal and the varying numbers of antennas at the BS generated the linear precoding problem, which could not be properly formulated. Under low power, MRT precoding maximized the SNR of each user targeted at the desired signal. The number of antenna arrays transmitting in the downlink (DL) determines the achievable high DR. By taking into consideration the comprehensive knowledge of large-scale fading and mitigated PC between adjacent cells, the transmitted data signals from BS build precoding paths to users.

[6-11] studied the impact of the pilot between neighboring cells by transmitting signals simultaneously, which provided a better estimation of the channel and improved pre-coding performance. Another author [12–17] looked into PC contractions using orthogonal pilots between adjacent cells and came up with the best PRS for the full user set. At the same time, the author [4], [18-22] studied the effects of PC and with the help of a pilot and without a pilot, and helped in a practical array MIMO to perform high DR. We propose in this research to employ PRS to reduce the PC of edge users by evaluating the channel with full knowledge of large-scale fading. Therefore, we evaluated the large-scale fading based on maximum ratio transmission (MRT) and zero forced (ZF) pre-encoding when the number of antenna elements and the number of users (UEs) increased by a large number and obtained a better performance analysis [23-25]. Massive MIMO systems have two major issues: pilot contamination and increased consumption power. Considering the evaluation of the proposed channel with full knowledge of large-scale fading, the PC reduction is capable of reaching high data speeds for pilot reuse. To generate a high-quality channel value, the BS links the training signal to a predefined PRS of each UE using comprehensive knowledge. Furthermore, the orthogonal PRS was used to eliminate PC in edge users whose channel quality began to deteriorate as the MRT and ZF pre-encoding approaches achieved large-scale performance and efficiency [26-30]. Increasing the high data rate of the network of 5G mobile systems depends on increasing the network capacity by evaluation channel. Because of the multicellular array's great degree of freedom and the enhanced pilot sequence for each neighboring cell, BS was exploited to evaluate channels. The number of users to share bandwidth has increased because of reduced PC. The pilot sequence provided by users of additional cells contaminated the limited coherent channel. To meet the ever-increasing demands of explosive wireless data services and prevent interference due to PC used in channel estimation across all BS, the number of PRS has not been arbitrarily increased. This coherence led to a critical limitation in channel estimation and reduced the possibility of data rates. In essence, PC results from assigning the same pilot sequences to users in nearby cells [31-35]. Based on the performance degradation of users who have not been assigned with the pilot sequence, the pilot assignment approach seeks to optimize the feasible sum rate for the target cell. Due to the restricted coherence time of the channel and the constrained bandwidth available in a multi-cell, the allocation of orthogonal pilot sequences for all users cannot be guaranteed. To improve the system's performance, a relative channel evaluation was used with full knowledge of the large-scale fading. Based on an evaluation of large-scale fading and efficacy study of MRT and ZF pre-encoding approaches, a PRS was used to cancel interference. These were pre-codes with less spatial dimensions that could block nearby cell interference from the most vulnerable UEs.

## II. SYSTEM MODEL

The base station (BS) delivers signals to the UEs in the following multicellular massive MIMO systems, where each cell has a BS M antenna to service $K$, ($M \gg K$). Where the channel model for the correlated Rayleigh fading channel matrix is independent and identically distributed (i.i.d.) and the properties minimum mean square error (MMSE) for channel $h_{ljk} \in \mathbb{C}^{M \times 1}$ can allocate a diverse antenna correlation to each channel between the users in the BS l and M in the BS $j$, the channel reciprocities were assumed to be the same in the uplink and downlink (DL). In $\Theta_{ljk} \in \mathbb{C}^M$, $M \times 1$ is the small-scale fading channel and $\Omega_{ljk} \in \mathbb{C}^{M \times M}$ accounts for the related channel correlation matrix for large-scale fading. In $[D_{lj}]_{k,k} = \sqrt{\Omega_{ljk}}$, $D_{ljk}$ represent the diagonal matrix and has a diagonal element which can be written as $\sqrt{\Omega_{lj}} = [\sqrt{\Omega_{lj1}}, .., \sqrt{\Omega_{ljK}}]$. The channel between BS $l$ and the $Kth$ user in cell j is given by

$$g_{ljk} = \sqrt{\Omega_{ljk}}\, \Theta_{ljk} \tag{1}$$

The BS was employed with imperfect channel state data (CSI). The $Kth$ user's established signal $z_{jk}$ inside the $jth$ cell can be written as

$$z_{jk} = \underbrace{\sqrt{\rho_d}g^H_{jjk}x_{jk}b_{jk}}_{desired\ signal} + \underbrace{\sqrt{\rho_d}\sum_{i=1,i\neq k}^K g^H_{jjk}x_{ji}b_{ji}}_{intra-cell\ interference} + \underbrace{\sqrt{\rho_d}\sum_{l=1,l\neq j}^L \sum_{i=1}^K g^H_{ljk}x_{lk}b_{lk}}_{inter-cell\ interference} + n_{jk} \tag{2}$$

where $g^H_{ljk}$ stands for the Hermitian transposition channel matrix, which is used to assess the UEs $K$ channel in each cell using an orthogonal pilot. $g_{jjk} = [g_{jj1} \ldots g_{jjK}] \in \mathbb{C}^{M \times K}$ is the DL channel between BS $j$ and user $K$ in its cell. The transmit signal vector of BS is $y_{lk} = x_{lk}b_{lk} \in \mathbb{C}^M$, $b_{lk} \in \mathbb{C}^{M \times K}$ is the linear precoding matrix, $x_{lk} \in \mathbb{C}^K \sim \mathcal{CN}(0, I_K)$ is the data transferred from cell $l$ to the users, $\rho_d$ represent the DL transmit power, and $n_{jk} \sim \mathcal{CN}(0_{M \times 1}, I_M)$ is the conventional noise vector.

### A. Channel Estimation

At the DL, we used channel estimate based on channel interactions in TDD. The main goal of this research was to improve the maximal DR with full knowledge of large fading to remove PC. By investigate the improved DR depends on using relative channel by reduce PRS to obtain the good quality of channel by used the training received signal with evaluated every pilot for the user. After that choosing, an optimal antenna selection depends on select the optimal a number of radio frequency chains (RF) and good quality of channel to exploit the high DR. We assumed that the reciprocity of the channel could be determined by matching the UEs in cell $j$ of the downlink $G^G_{jjk}$, which is the Hermitian transposition of the uplink $G^G_{jjk}$. The channel might be estimated using preparation signal in the same identical PRS for neighboring cells as

$$A_{jk} = g_{jjk} + \sum_{l\neq j}^K g_{ljk} + \frac{n_{jk}}{\rho^{1/2}} \tag{3}$$

where $\rho$ is the transmit power, which is relative to the current pilot SNR, and is the length of the training transmit power.

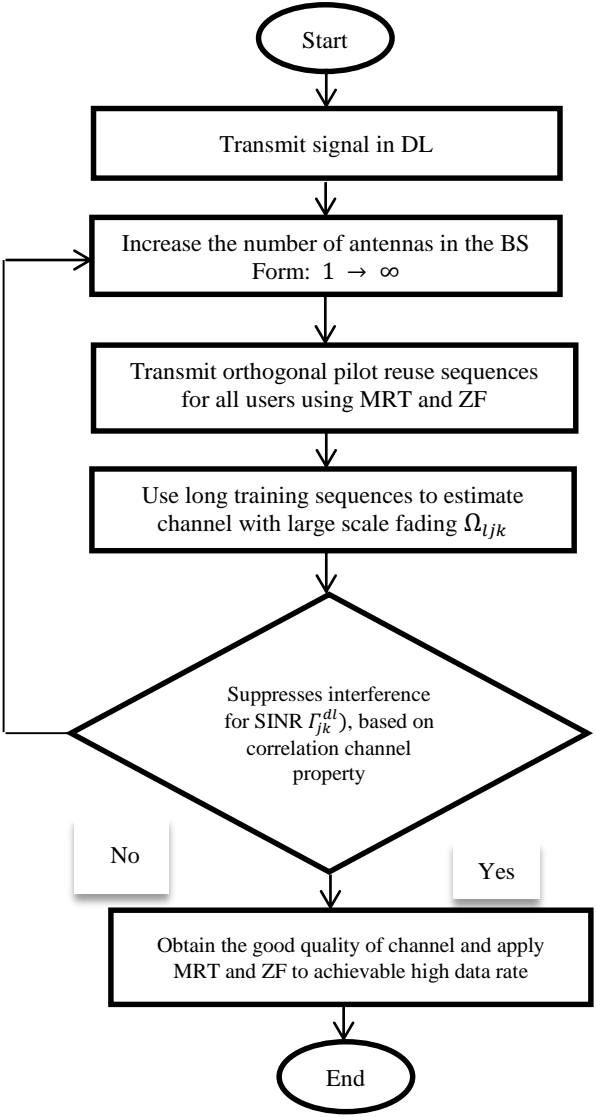

Fig. 1. Flow chart of achievable high data rate.

$$\hat{g}_{jjk} = \Omega_{jjk}\varphi_{jk} \sum_{l \neq j}^{K} g_{ljk} + \frac{n_{jk}}{\rho^{1/2}} \quad (4)$$

*B. Downlink Transmission*

In this section, a large-scale fading, PC was suppressed by increasing the number of transmit pilot reuses. The achievable high data rate derived by using linear combining by propose relative channel estimation in full knowledge of large-scale fading. Then, the downlink derived the achievable DR and SINR based on linear precoding. Consequently, mitigated pilot contamination allowed an increasing number of users to share bandwidth and improved the system performance to get better channel estimation and reducing inter-cell interference. The relative channel estimation is very important to evaluate the conventional pilot and suppresses interference between neighboring cells. Based on only utilized large-scale fading, the achievable data rate was enhanced, assuming that the BS had imperfect CSI. In addition, the interference between users was suppressed without using more time-frequency resource. The BS transmitted the pilot sequences to every UEs in the cell. These UEs estimated their own channels.

Furthermore, symbols $K \times L$ were required for the orthogonal PRS in the multi-cell. As a result, extensive training channel were utilized, with the variance channel expressed as

$$\varphi_{jk} = \left(\frac{I_M}{\rho} + \sum_{l=1}^{L} \Omega_{ljk}\right)^{-1} \quad (5)$$

A pilot sequence of $Y_p = K = \infty$ was created by increasing the orthogonal PRS, where $Y_p$ is the symbol of the DL pilot sequence that performed for the desired channel; the received signal is provided by

$$\xi_{ljk} = \sqrt{\rho_d Y_p} \sum_{l=1}^{L} g_{ljk} + n_{lk} \quad (6)$$

For the estimation channel, the properties of MMSE $\tilde{g}_{ljk} = \hat{g}_{jlk} - g_{ljk}$ the channel estimation between distributed more users $K$ in same cells $L$ was discovered, where $g_{jjk} \sim \mathcal{CN}(0, \varphi_{jk})$ is the uncorrelated estimation self-regulating of $\tilde{g}_{jjk} \sim \mathcal{CN}(0, (\varphi_{jk} - O_{ljk})I_M)$, and $O_{ljk} = \Omega_{jjk}\varphi_{jk}\Omega_{ljk}$, which is also self-governing of antenna $M$ for large-scale fading [6-10],[22-28]. Enhancing quality of channel for $Kth$ UEs in terms of MMSE can be written as

$$\tilde{g}_{ljk} = \mathbb{E}\|\hat{g}_{ljk} - g_{ljk}\|^2$$
$$= \mathbb{E}\{|\varphi_{jk}(\xi_{ljk} +) - g_{ljk}|^2\}$$
$$= \mathbb{E}\{|(\varphi_{jk} - I_M)(\xi_{ljk}) + n_{lk}|^2\}$$
$$= \{\Omega_{ljk}(\varphi_{jk} - I_M)(\varphi_{jk} - I_M)^G(\xi_{ljk}) + \sigma^2 I_M\}$$
$$\tilde{g}_{ljk} = \{\Omega_{ljk}(\varphi_{jk} - I_M)(\varphi_{jk} - I_M)^G(\Theta_{ljk}(\rho_d Y_p \sum_{l=1}^{L} \Omega_{ljk})\Theta_{ljk}^G)\xi_{ljk} + \sigma^2 I_M\}$$
$$\tilde{g}_{ljk} = \{\Omega_{ljk}((\frac{I_M}{\rho} + \sum_{l=1}^{L} \Omega_{ljk})^{-1} - I_M)((\frac{I_M}{\rho} + \sum_{l=1}^{L} \Omega_{ljk})^{-1} - I_M)^G((\Theta_{ljk}(\rho_d Y_p \sum_{l=1}^{L} \Omega_{ljk})\Theta_{ljk}^G)\xi_{ljk}) + \sigma^2 I_M\} \quad (7)$$

The good quality of channel can be obtained by applied MMSE and also determined by $\xi_{ljk}$ by employing matrix inversion with $\Theta_{ljk}\Theta^G_{ljk} = 1$ and the same vector $\Theta^G_{ljk}\xi_{ljk}$ is proportionate to the characteristics of MMSE. The estimation channel is proportional based on the training received signal $\Theta^G \xi_{ljk}$ is

$$\tilde{g}^G_{ljk} / \|\tilde{g}_{ljk}\| = \frac{\Theta^G_{ljk}\xi_{ljk}}{\|\Theta^G_{ljk}\xi_{ljk}\|} \quad (8)$$

Reducing PC in (8) depend on when the $\mathbb{E}[g_{ljk}a_{lk}] = \mathbb{E}\|\tilde{g}^G_{ljk}\|$. From (4) and (8), the quality of channel $\tilde{g}_{ljk}$ ionly connected to channel response $g_{ljk}$, which may be represented as

$$\tilde{g}_{ljk} = \frac{\rho_d Y_p \Omega_{ljk}}{1 + \rho_d Y_p \sum_{i=1}^{L} \Omega_{lik}} \Theta^G_{ljk}\xi_{ljk} \quad (9)$$

*C. Achievable Data Rate*

In this section, to remove PC in edge UEs with reduced channel quality of large-scale fading and evaluate MRT and ZF precoding. The study achieved lower bounds on the DL for DR and SINR. This was performed by allocating PRS to a user group that alleviated PC when the number of antenna $M$ increased to infinity. The training signal at the BS cause the linear precoding problem, which cannot be properly expressed [29-33]. In addition to the MRT, the MRT precoding maximised the SNR of each UE $K$ for the required signal under allocated power, due to interference of $\mathcal{A}_j = [b_{1,MRT}, ., b_{K,MRT}] = [\tilde{g}^G_{lj1}, ., \tilde{g}^G_{ljk}] = \tilde{G}^G_{ljk}$. Precoding with ZF able to deliver data, and improve processing to account for background noise based on the situation of channel $b_j = \tilde{G}_{ljk}(\tilde{G}^G_{ljk}\tilde{G}_{ljk})^{-1} / \left\| \tilde{H}_{ljk}(\tilde{G}^G_{ljk}\tilde{G}_{ljk})^{-1} \right\|^2$. To receive the lower bound of DR as:

$$\mathcal{R}_{jk} = \sum_{l=1}^{L}\sum_{i=1}^{K}(1-\partial)\left[\log_2(1+\Gamma_{jk}^{dl})\right] \qquad (10)$$

where $(1-\partial)$ represent the loss of pilot signalling for the pre-log factor, and $0 < \partial < 1$ indicates the possible DR. A correlated channel matrix was used to predict the channel response based on mitigated PC. We rotten the received signal as

$$\mathcal{Y}_{jk} = \sqrt{\rho_d \Upsilon_p}\mathbb{E}\{g^G_{jjk}b_{jk}\}x_{jk} + \sqrt{\rho_d \Upsilon_p}\sum_{i=1,i\neq k}^{K}(g^G_{jjk}b_{jk} - \mathbb{E}\{g^G_{jjk}b_{jk}\})\mathcal{V}_{jk} + \sqrt{\rho_d \Upsilon_p}\sum_{l=1,l\neq j}^{L}\sum_{i=1}^{K}g^G_{ljk}b_{lk}\,x_{lk} + n_{jk} \qquad (11)$$

From (11) the uncorrelated noise channel it be the worst case. To evaluated the achievable data rate at transmit signal from BS, which can be written as

$$R_t = \sum_{l=1}^{L}\sum_{i=1}^{K}(1-\partial)\log_2\left(1 + \frac{\mathbb{E}|DS|^2}{\mathbb{E}|Uncorrelated\ noise|^2}\right) \qquad (12)$$

The PRS was found to be linked between the channel $h^H_{ljk}$ and the neighbouring cell's precoding $b_{lk}$. The SINR was calculated at the $kth$ UEs[32-34], and the desired SINR signal can be written as

$$\mathbb{E}|DS|^2 = \rho_d \Upsilon_p \mathbb{E}\left[g^H_{jjk}b_{jk}\right]^2 \qquad (13)$$

Power can be calculated from uncorrelated noise as
$\mathbb{E}|Uncorrelated\ noise|^2 =$
$\rho_d \Upsilon_p \sum_{l=1,l\neq j}^{L}\sum_{i=1}^{K}\mathbb{E}\left[\left|g^G_{ljk}b_{lk}\right|^2\right] -$
$\sum_{i=l}^{L}\rho_d \Upsilon_p \left|\mathbb{E}\left[g^G_{jjk}b_{jk}\right]\right|^2 + \sigma^2 \qquad (14)$

Under derived both MRT and ZF precoding to produce remarkable SINR outcomes with huge numbers of $M$.

$$\Gamma_{jk}^{dl} = \frac{\rho_d \Upsilon_p\left|\mathbb{E}\left[g^G_{jjk}b_{jk}\right]\right|^2}{\rho_d \Upsilon_p \sum_{l=1,l\neq j}^{L}\sum_{i=1}^{K}\mathbb{E}\left[\left|g^G_{ljk}b_{lk}\right|^2\right] - \sum_{i=l}^{L}\rho_d \Upsilon_p \left|\mathbb{E}\left[g^G_{jjk}b_{jk}\right]\right|^2 + \sigma^2} \qquad (15)$$

We calculated the interference channel and power allocation in terms of MRT precoding, which can be represented as

$$\rho_d\left|\mathbb{E}\left[g^G_{ljk}b_{lk}\right]\right|^2 = \frac{d}{M\,var(g_{ljk})}\left|\left\{\mathbb{E}\left\|\tilde{g}_{ljk}\right\|^2\right\}\right|^2 \qquad (16)$$

The considerable fading resulted in the noise variation. The covariance channels matrices were used $\left|\mathbb{E}\left[g^G_{ljk}b_{lk}\right]\right|^2$ from Kay [6]: $\rho_d\left|\mathbb{E}\left[g^G_{ljk}b_{lk}\right]\right|^2 = Mvar(g_{ljk})$ [32-34]. By employing correlated channels in the same cell to adjacent cells in (14) the impact of PC for interference can be mitigated (9). For MRT, the SINR can be written as

$$\Gamma_{jk}^{dl-mrt} = \frac{Mvar(\tilde{g}_{jjk})}{Mvar(\tilde{g}_{ljk}) + \sum_{l=1,l\neq j}^{L}\sum_{i=1}^{K}var(\tilde{g}_{jjk}) + \frac{\sigma^2}{\rho_d}} \qquad (17)$$

Based on ZF precoding the interference between neighboring cell expressed as $b_{lk} = \mathbb{E}\left|\tilde{G}_{ljk}(\tilde{G}^G_{ljk}\tilde{G}_{ljk})^{-1}\right| / \mathbb{E}\left\{\left\|\tilde{G}_{ljk}(\tilde{G}^G_{ljk}\tilde{G}_{ljk})^{-1}\right\|^2\right\}^{1/2}$, the inter-user interference reduced as $b_{lk} = (M-K)var(\tilde{g}_{ljk})^{1/2}G_{ljk}(\tilde{G}^G_{ljk}\tilde{G}_{ljk})^{-1}$. Subject on a number of $M \to \infty$, the variance channel is $var\{\tilde{g}^G_{ljk}b_{jk}\}\xrightarrow[M\to\infty]{} 0$. The number of PRS must be established in order to attain the high performance of the achievable DR. In order to get improved performance, the channel estimation must have a high SINR. From (15), the linear precoding of ZF can be written as

$$\rho_d\left|\mathbb{E}\left[g^H_{ljk}b_{lk}\right]\right|^2 = \rho_{jk}(M-K)\,var(\tilde{g}_{ljk}) \qquad (18)$$

The interference can be written in the denominator as

$$\rho_d \sum_{l=1,l\neq j}^{L}\sum_{i=1}^{K}\mathbb{E}\left[\left|g^H_{ljk}b_{lk}\right|^2\right] - \rho_{jk}\left|\mathbb{E}\left[g^H_{jk}b_{jk}\right]\right|^2 + \sigma^2 = \rho_d(M-K)\,var(\tilde{g}_{ljk}) + \sum_{l=1,l\neq j}^{L}\sum_{i=1}^{K}\rho_d var(\tilde{g}_{ljk}) + \sigma^2 \qquad (19)$$

At employed ZF precoding, the large scale fading channel is obtained by replacing (18) and (19) into (17).

$$\Gamma_{jk}^{dl-zf} = \frac{(M-K)var(\tilde{g}_{jjk})}{(M-K)var(\tilde{g}_{ljk}) + \sum_{l=1,l=j}^{L}0_{jk} + \sum_{l=1,l\neq j}^{L}\sum_{i=1}^{K}var(\tilde{g}_{jjk}) + \frac{\sigma^2}{\rho_d}} \qquad (20)$$

### D. Our Proposed

According to the random distribution of $K$ in each cell, the SINR depend on the quality of channel fading $\Omega_{ljk}$ in equation (10). Because the training signal at the BS cannot be directly constructed, MRT and ZF linear precoding must be utilized to improve multi-cell interference in MIMO systems without adding time. The performance evaluation of MRT and ZF precoding techniques was achieved when the number of antenna elements M and number of UEs K increased to large numbers. The perfect orthogonality PRS, focusing on the impact of pilot contamination and the analysis of MRT and ZF precoders with PC in DL multi-cell massive MIMO systems. The mitigated PC allows an increasing number of users to share bandwidth in order to improve the system performance to get better channel estimation and reducing inter-cell interference. The relative channel estimation is very important to evaluate the conventional pilot and suppresses interference between neighboring cells. This is because the channel estimation was more efficient with full knowledge of large-scale fading. In addition, orthogonality at its best with the help of PRS, the quantity of UEs can be improved by reducing cross-tier interference across several antennas. When transmission power is limited in mm-wave, the perfect orthogonality PRS is crucial for increasing capacity and high coverage at cell edges, lowering high path loss, and suppressing cross-tier interference. The desired trade-off between improving channel estimation accuracy and retaining more resources for data transmission was achieved because of the improvement of PRS. Suppressing PC to achieve high DR for pilot reuse by considering the proposed channel estimation with comprehensive knowledge of large-scale fading. The comprehensive knowledge this means the BS correlates the training signal with the established PRS of every UE to obtain a high-quality channel estimate. In addition, the orthogonal PRS applied to remove PC in edge UEs with reduced channel quality according to the approximation of large-scale fading and performance evaluation of MRT and ZF precoding techniques.

As a result, the best quality of channel $\Omega_{ljk}$ of UEs in $ith$ cells depend on sorting UEs when $j \neq l$. The channel quality of UEs increased by transmission orthogonal PRS to the edge UE group in equations (17) and (20) based on quality of channel $\Omega_{jlk}$ and separating UEs into two groups as follows:

$$\Omega_{ljk} = \ell_i \geq \tau\mu_i \rightarrow \begin{cases} Yes \rightarrow center\ users \\ No \rightarrow edge\ users \end{cases} \quad (21)$$

where $\tau\mu_i$ is the threshold value of grouping, and $\ell_i$ represent the channel quality of UEs $\Omega_{ljk}$, the DR enhanced based on selected the quality of channel of the edge UEs. As a result, acceptable channel quality can be achieved by allocating orthogonal PRS to the edge users while reusing the same PRS to the center users [12], [19], [28], $\mu_i$ a user grouping that may be expressed as

$$\mu_i = \sum_{k=1}^{K} \frac{\max[\ell_{i1}, \ell_{i2}, \ldots, \ell_{iK}] + \min[\ell_{i1}, \ell_{i2}, \ldots, \ell_{iK}]}{2} \quad (22)$$

where $K_{ic} = \text{card}[k:\ldots, \ell_i > \mu_i]$ represents the number of center UEs and $K_{ie} = \text{card}[k:\ldots, \ell_i \leq \mu_i]$ represents the number of edge UEs. The perfect orthogonality PRS, with an emphasis on the impact of PC and an investigation of MRT and ZF precoders with PC. The mitigated PC lets a growing number of users to share bandwidth, improving system performance and eliminating inter-cell interference. The perfect orthogonality PRS can reduce the cross-tier interference between multiple antennas to assist the number of UEs [35-40], the high DR achieved in terms of MRT and ZF by mitigated PC as follows:

$$\mathcal{R}^{mrt,\ zf}_{jk} = \sum_{l=1}^{L} \sum_{i=1}^{K} \left(1 - \frac{\partial}{K}(\sum_{l=1}^{L} LK_{ie} + \max[K_{ic}, K_{ic}, \ldots, K_{ic}])\right) \log_2(1 + \Gamma_{jk}^{dl-mrt,zf}) \quad (23)$$

Large-scale fading caused some losses in the DR in equation (23), which can be prevented by growing the number of $M$ and confirming the location UEs $K$.

## III. SIMULATION RESULTS

Figure 2 shows how a constant number of UEs inside each cell and quality of channel offered complete information of large-scale fading at greater spatial resolutions. When the grouping value was increased to M = 256 and K = 10, the target cell achieved a high DR. Figure 2 also demonstrates that increasing the PRS from 1 to 3 enhanced the DR that could be achieved. Furthermore, because the ZF was prepared to operate at high SINR, it gave greater DR than MRT for big pilot reuse. Furthermore, by adjusting the number of edge users or the grouping parameter, the DR that can be achieved can be increased. Finally, the multi-increased cell's antenna count rendered these channels orthogonal to others, reducing interference between adjacent cells. In addition, when compared to MRT, ZF precoding a mitigate interference resulting in a high attainable DR. The average DR for all curves was nonlinear to the number of UEs, as shown in Fig. 2, necessitating the selection of the optimal number of K in order to achieve a high DR. Furthermore, a high DR was attained based on the impacts of bandwidth and correlation between the transmit pilots. Furthermore, the user's capacity is limited, limiting the transmit orthogonal PRS based on bandwidth impacts. With more transmitting PRS in large-scale fading, the channel estimation evaluated at the received signal and suppressed PC was dependent. Based on only utilized large-scale fading, the achievable data rate was enhanced, if the BS had imperfect CSI. In addition, the interference between users was suppressed without using more time-frequency resource. The BS transmitted the pilot sequences to every UEs in the cell. These UEs estimated their own channels. The achievable DR was improved and established based on added antennas to existing cell. Using different PRS in DL reduced the performance loss and provided better estimation quality in the channel, maximizing high DR.

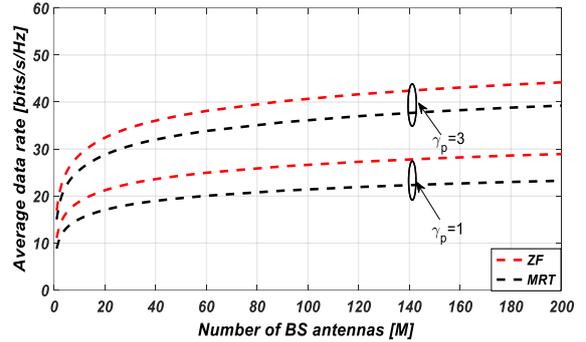

Fig. 2. Data rate versus a number of antennas M

According to equation (23), a high feasible DR may be attained, based on UE grouping, to reduce PC in edge UEs. Furthermore, when the quantity of $\Upsilon_p = 7$ was used, the system performance resulted in a significant DR performance as compared to $\Upsilon_p = 1$. Figure 3 demonstrates that the number of PRS did not grow arbitrarily in order to reduce interference due to the PC used in every BS for form channel estimation, which resulted in a major restriction in coherence channel estimation and a reduction in DR capacity. The average DR began to rise and then progressively fell in Fig. 3, indicating that the increasing DR of edge UEs is greater than the decreasing DR of center UEs.

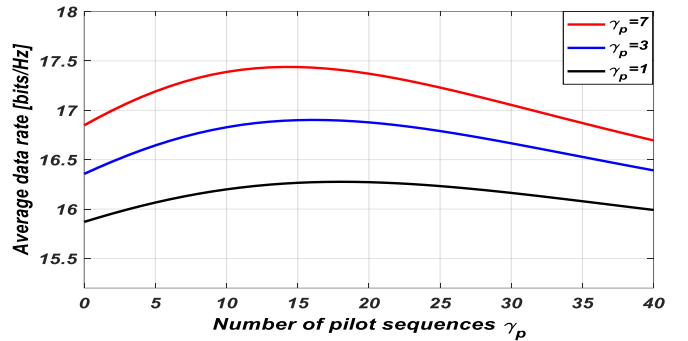

Fig.3. High data rate versus number of pilot sequences $\Upsilon_p$.

As seen in Fig. 4, increasing a number of cells it reducing the high DR due to interference. Interference happened when the number of $M$ was less than the number of cells $M < L$, since users of different cells utilised the same pilot sequence, contaminating the channel estimation. To limit the interference caused by pilot contamination, which generated channel estimation at all BSs, the number of pilot reuse sequences could not be increased arbitrarily. This resulted in a major constraint in the calculation of coherence channels, lowering the data rate's capacity. Because of the channel coherence interval limitation, the orthogonal PRS cannot be allocated to all UEs in each cell. The short coherence intervals able to reduce the PC and improve the quality of channel. Furthermore, as the number of cells increased, the average data rate declined as the number of pilot reuses increased. The average data rates are R= (55.5, 50) bits/s/Hz when the number of cells was L=2 and PRS was $Y_p$ =7. Meanwhile, due to the essential constraint in coherence channels estimation, the average DR fell at $\mathcal{R}$ = (28.6, 26.2) {bits/s/Hz}, with the higher number of cells L=18 and the same PRS.

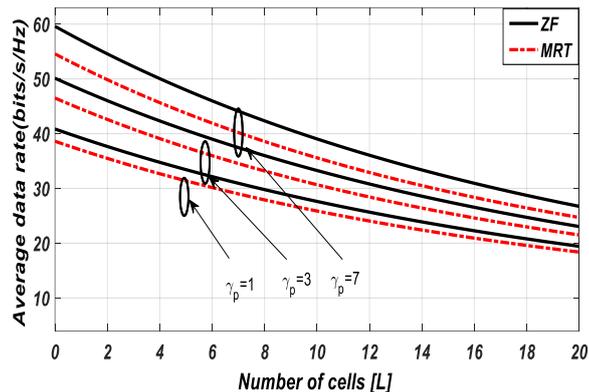

Fig. 4. Data rate versus a number of cells L.

## IV. CONCLUSION

The evaluation of MRT and ZF precoders with PC in a DL multi-cell massive MIMO system was reported in this paper. The good coverage channel is very important based on recognized PC connected with good, received signal to UEs due to channel estimation. For the edge users in surrounding cells with large-scale fading, orthogonal PRS was used to evaluate channel estimate accuracy to eliminate PC. The numerical findings revealed that using orthogonal PRS in the DL decreased loss, improved channel estimate quality, and increased data throughput. In future works, we will study the efficient pilot and channel scheduling for small-scale fading.

ACKNOWLEDGMENT

This work was supported by the Ministry of Higher Education Malaysia through the Fundamental Research Grant Scheme FRGS/1/2019/TK04/UTHM/02/8 and Universiti Tun Hussein Onn Malaysia.